\documentclass{emulateapj}
\usepackage{apjfonts}

\begin{document}

\title{Recognizing the First Radiation Sources Through Their 21--cm Signature}
\author{Leonid Chuzhoy$^{1}$, Marcelo A. Alvarez$^{1}$ and Paul R. Shapiro$^{1}$}
\bigskip

\altaffiltext{1}{McDonald Observatory and Department of Astronomy, The
University of Texas at Austin, RLM 16.206, Austin, TX 78712, USA;
chuzhoy@astro.as.utexas.edu; marcelo@astro.as.utexas.edu;
shapiro@astro.as.utexas.edu} 

\begin{abstract}
At the beginning of the reionization epoch, radiation sources produce
fluctuations in the redshifted 21-cm background. We show that
different types of sources (such as miniquasars, Pop II and III stars,
supernovae, etc.) produce distinct signatures in the 21-cm signal radial profiles and statistical fluctuations,
through which they can be identified. Further, we show that the 21-cm
signal from X-ray emitting sources is much easier to observe than was
expected, due to a previously neglected pumping mechanism. 
\end{abstract}


\section{\label{Int}Introduction}

As shown by WMAP observations \cite{Sp}, within a half billion years
after the Big Bang the Universe must have given birth to numerous
radiation sources, that by $z\sim 10$ reionized most of the gas in the
intergalactic medium (IGM).   However, the nature of these sources
remains a puzzle. Miniquasars, Pop II and III stars, supernovae, and
many others have been suggested as the main engines of reionization \cite{CF}.
Unfortunately, direct observations may not be able to distinguish
between the different types. For most, the low-energy part of the
spectrum has the same thermal shape, $J(\nu)\propto \nu^2$, while the
upper part, $h\nu>10.2$ eV is absorbed or scattered by the neutral
IGM. However, as we show in this paper, it is still possible to
measure the shape of the upper part of the spectrum, from the source
imprint on the 21-cm signal of the surrounding neutral IGM.

When at redshift $z$ the hydrogen spin temperature, $T_s$,
is different from the temperature of the cosmic microwave background
(CMB), $T_{\rm CMB}$, the CMB brightness temperature presently
observed at the wavelength $21(1+z)$ cm changes by 
\begin{eqnarray}
\delta T_{\rm b}=0.03\; {\rm K} \left(\frac{T_{\rm s}-T_{\rm CMB}}{T_{\rm s}}\right)(1+\delta)(1-x)\times  \nonumber \\
\left(\frac{\Omega_{b0} h_0}{0.03}\right) \left(\frac{\Omega_{m0}}{0.25}\right)^{-1/2}  \left(\frac{1+z}{10}\right)^{1/2}.
\end{eqnarray}
where $\delta$ is the local overdensity and $x$ is the ionized hydrogen fraction.
The spin temperature is determined by \cite{f8} 
\begin{eqnarray}
\label{Tsp}
T_{\rm s}=\frac{T_{\rm CMB}+y_\alpha T_\alpha+y_{\rm c}T_{\rm k}}{1+y_\alpha+y_{\rm c}},
\end{eqnarray}
where $T_{\rm k}$ is the kinetic temperature, $y_{\rm c}$ is a
constant proportional to the collisional excitation rate  (see
Zygelman (2005) and Liszt (2001) for the contribution of neutral atoms
and electrons to the collisional pumping) and $y_\alpha$ is given by 
\begin{eqnarray}
y_{\alpha}=\frac{16\pi^2 T_*e^2 f_{12}J_0}{27 A_{10}T_{\rm k} m_{\rm e}c}S_\alpha,
\end{eqnarray}
where $f_{12}=0.416$ is the oscillator strength of the Ly$\alpha$
transition, $A_{10}$ is the spontaneous emission coefficient of the
hyperfine transition,  $J_0$ is the radiation intensity at the far red
wing of the Ly$\alpha$ resonance (where the intensity of UV photons is
unaffected by scatterings) and $S_\alpha$ is a correction factor,
which for the unperturbed IGM is given by \cite{CS} 
\begin{eqnarray}
\label{cor}
S_\alpha=e^{-0.37(1+z)^{1/2}T_{\rm k}^{-2/3}}\left(1+\frac{0.4}{T_{\rm k}}\right)^{-1}.
\end{eqnarray}
Several authors have previously explored the ways $T_s$ can be
decoupled from $T_{\rm CMB}$ by the first radiation sources.  Nusser
(2005) has pointed out that, since charged particles have a much
higher cross-section for collisionally de-exciting the hyperfine
transition, $T_s$  can be decoupled from $T_{\rm CMB}$ in moderately
overdense partially ionized regions (which may be produced by an
X-ray source or following the recombination of a fully ionized HII
region). In this way a miniquasar with luminosity $L=10^{41}\; {\rm
erg\cdot s^{-1}}$ can decouple the gas within a few hundred comoving
kiloparsec \cite{KMM}. Cen (2005) considered a massive starburst in
the first galaxies that produce a strong UV background. In his model
$T_s$ is decoupled from   $T_{\rm CMB}$ by the photons slightly on the
blue side of the Ly$\alpha$ resonance that redshift into the resonance
within some distance (typically up to a few comoving Mpc) from the
source. However, in this paper we show that in all of the above
systems the spin temperature is dominated by locally emitted
Ly$\alpha$ photons that can be produced in three different ways
(collisional excitation, recombination, or cascade). Including these
photons in the calculation not only boosts the intensity of the signal
(by orders of magnitude for the X-ray sources), but makes a
qualitative change in its radial profile.

In \S 2, 3 and 4 we recalculate the 21 cm signal from the systems
discussed above (X-ray and UV sources, and recombining HII regions).
In \S 5 and \S 6  we consider the signal produced by galactic winds
and short-lived radiation sources, respectively. In \S 7 we summarize
our results. 

\section{X-ray sources}
After traveling an average comoving distance of $\lambda(\nu)\approx 0.1\;
{\rm Mpc}\;((1+z)/20)^{-2}(h\nu/0.1\;{\rm keV})^3$ through the neutral
gas, an X-ray photon gets absorbed by an atom. A high energy electron
produced by the absorption scatters with other atoms, electrons, and
protons, thereby depositing its energy into heat, secondary
ionizations, and atomic excitations.  Shull \& Van Steenberg (1985)
showed that the division of energy depends on the ionized fraction
$x$, with $\phi_H \approx [1-(1-x^{0.27})^{1.3}]$ of the total going
to heat, $\phi_i\approx 0.39(1-x^{0.4})^{1.76}$ to ionizations and
$\phi_\alpha\approx 0.48(1-x^{0.27})^{1.5}$ to excitations of hydrogen
atoms. 

On the basis of these results we find that when the initial ionized fraction
is low (as it is for the primordial gas after $z\sim 1000$) and
temperature is below $10^4$ K, the X-rays increase $T_{\rm k}$ and $x$
by  
\begin{eqnarray}
 \frac{\Delta T}{1\; {\rm K}}\approx 3\cdot 10^3 \left(\frac{\epsilon}{1\; {\rm eV}}\right)^{5/4}, \; \nonumber \\
\Delta x\approx 6\cdot 10^{-5}  \left( \frac{\Delta T}{1\; {\rm K}}\right)^{3/4},
\end{eqnarray}
where $\epsilon$ is the average energy deposited per atom.

The heating rate and intensity of Ly$\alpha$ photons produced by
deexcitations of hydrogen atoms, at a distance $R$ from the X-ray
source, is given by 
\begin{eqnarray}
\label{dT}
\frac{3}{2}n_H k\frac{dT_k}{dt}&=&\phi_H(x) \int \frac{L_\nu e^{-R/\lambda(\nu)}}{4\pi R^2\lambda(\nu)} d\nu \\
\label{Ja}
J_0&=&\frac{\phi_\alpha(x)c}{4\pi H\nu_\alpha} \int \frac{L_\nu e^{-R/\lambda(\nu)}}{4\pi R^2\lambda(\nu)} d\nu,
\end{eqnarray}
where $L_\nu$ is the source luminosity at frequency $\nu$ and $n_H$ is
the hydrogen atom number density.  

Assuming that the spectrum of high-redshift X-ray sources is
similar to "ultraluminous" X-ray sources in nearby galaxies,
$L_\nu\approx L_0 \nu^{-1}$ for $0.1<h\nu<2$ keV \cite{Mil}, we
can transform eqs. \ref{dT} and \ref{Ja} into  
\begin{eqnarray}
\label{dT2}
\Delta T&\approx& 0.5\; {\rm K} \left(\frac{L_0}{10^{40}\; {\rm ergs \;s^{-1}}}\right)^{5/4} \left(\frac{R/(1+z)}{1\; {\rm Mpc}}\right)^{-15/4}t^{5/4} ,\\
J_0&\approx& \phi_\alpha  \left(6\cdot 10^{-5}  (\Delta T/{1\; {\rm K}})^{3/4}\right) \left(\frac{L_0 c}{48\pi^2 R^3 H \nu_\alpha}\right),
\end{eqnarray}
where $t$ is given in Myrs.

Figure 1a shows the evolution of 21-cm brightness temperature. Close to the source, where the gas is heated above $T_{CMB}$, 
$T_b$ is positive. Farther out the gas becomes colder, so that $T_b$ becomes negative. Still farther out, the pumping radiation 
becomes too weak to decouple $T_s$ from $T_{CMB}$, and  $T_b$ approaches a small 
limiting value that is given by collisional coupling only.

\begin{figure}
\resizebox{\columnwidth}{!}
{\includegraphics{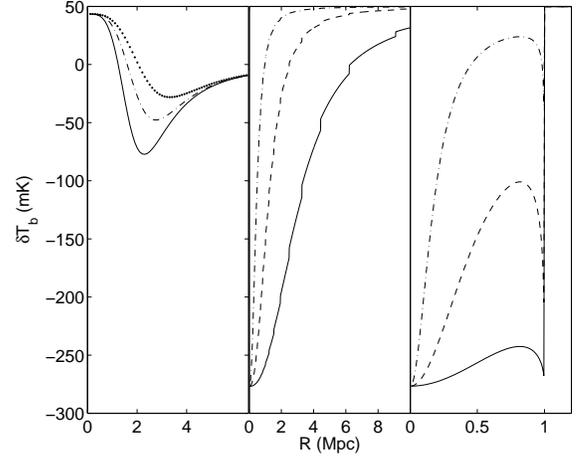}}
\caption{\label{Lym}The brightness temperature versus comoving radius around a radiation source at $z=20$. a) X-ray source with $L=10^{42}\; {\rm erg/s}$ and the age of $t=1$ (solid), 2 (dash-dotted)  and 4 Myr (dotted line). b) UV source with total luminosity of, respectively, $L=5\times 10^{41}$ (dashed), $5\times 10^{42}$ (dash-dotted) and $5\times 10^{43} {\rm erg/s}$ (solid line) between Ly$\alpha$ and Ly-limit frequencies. c) Galactic wind with $L_\alpha=10^{41}$ (solid), $10^{40}$ (dashed) and $10^{39}{\rm erg/s}$ (dash-dotted line). } 
\end{figure}

\begin{figure}[b]
\resizebox{\columnwidth}{!}
{\includegraphics{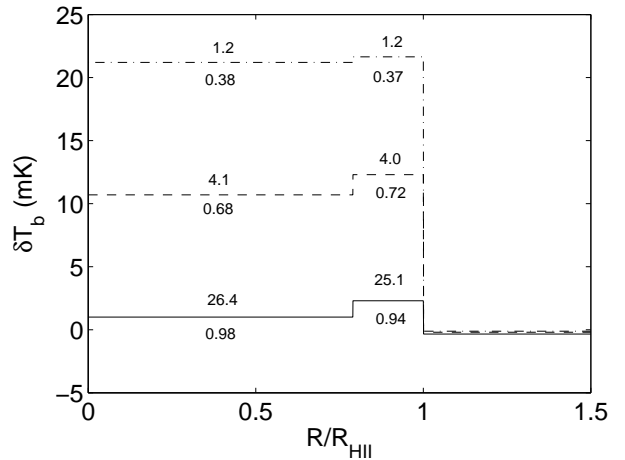}}
\caption{\label{HII1} The brightness temperature versus comoving
radius of the recombining HII region that was ionized at $z=20$. The
solid, dashed and dash-dotted lines correspond to $z=19$, 17 and 15. The numbers above and below the lines 
show, respectively, $T_k/1000$ K and $x$ at a given redshift.}
\end{figure}

\section{UV sources}

The decoupling of the spin temperature may also be done by radiation
sources that emit a significant fraction of their energy in the  range
between Ly$\alpha$ and Ly-limit frequencies.  Unlike ionizing photons,
photons in this range can travel freely through neutral hydrogen,
until they are  redshifted into one of the atomic resonances.  With
the exception of photons that are very close to the Ly-limit, which
may skip one or more resonances, most of these photons are  first
scattered when the closest resonance is reached. Thus, only photons
emitted between Ly$\alpha$ and Ly$\beta$ frequencies reach the
Ly$\alpha$ resonance, which they leave only after $\sim 10^6$
scatterings. By contrast, photons emitted above the Ly$\beta$
frequency  typically scatter only a few times before being destroyed
by cascading to lower levels. A fraction of resonant photons destroyed
by cascade (0 for Ly$\beta$, 0.26  for Ly$\delta$ and  $\sim 1/3$ for
higher resonances) produce Ly$\alpha$ photons as one of the cascade
products \cite{Hir,PF}. Since the frequency  range between the Ly$\gamma$ and
Ly-limit is $2.2$ times smaller than that between Ly$\alpha$ and Ly$\beta$,
Ly$\alpha$ photons produced directly through redshifting are
generally more numerous than those produced by cascade. However, the
latter photons are typically  produced within a much shorter distance
from the source (on average the photon between $n$ and $n+1$ resonance
travels a distance of  $\sim n^{-3}H^{-1}c/2$ before reaching a
resonance), which makes them a dominant pumping mechanism within a few
Mpc from the source.  Including these photons in the calculation not
only boosts the pumping radiation intensity, but also changes its
radial profile.  Whereas normally the radiation intensity is
proportional to $R^{-2}$, the cascades change this dependence to
approximately $R^{-7/3}$.  In addition, they also result in the
discontinuous jumps in pumping radiation intensity and 21 cm
brightness signal at  $R_n=(n^{-2}-(n+1)^{-2})H^{-1}c$. Figure 1b shows the radial profile of 21-cm brightness temperature 
of the unheated IGM around UV sources. In principle, the resonant scatterings affect the gas temperature, which 
therefore becomes time- and position-dependent \cite{CS6}. However, unlike the X-ray photons, the UV resonance photons generally 
gain or lose only a very small fraction of their energy in exchange with the gas, so in Figure 1b we neglect this energy transfer.

\section{Recombining HII regions}

The low-mass galaxies that are first to form in the standard
$\Lambda$CDM model are expected to be very vulnerable to
feedback. During brief and intensive periods of star formation, such
galaxies may be completely disrupted by SN explosions or
photoevaporation.  Furthermore, reionization may also delay gas
accretion and star formation in the neigbouring halos. Consequently,
the H II regions  produced by the first galaxies may be given a pause,
allowing them to cool and recombine.

\begin{figure}[b]
\resizebox{\columnwidth}{!}
{\includegraphics{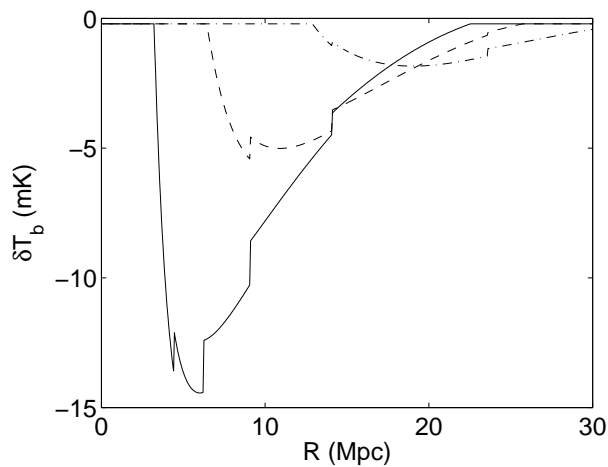}}
\caption{\label{Lymtd} The brightness temperature versus comoving radius around a short-lived UV source at $z=20$. The intensity of the source is assumed to vary as $L=5\times 10^{43} {\rm erg/s}\times \sin(\pi t/3 Myr)$ for  $t<3$ Myr. The solid, dashed and dash-dotted lines correspond to  $t=3.5$, 4 and 5 Myr.}
\end{figure}

Unless the gas is extremely overdense, its
thermal evolution as it recombines is determined mainly by inverse 
Compton cooling and radiative recombination. 
Right after ionization the temperature of the gas in the IGM depends
on the shape of the ionizing spectrum. If the shape changes
appreciably during the life-time of the ionizing source or if the
source life-time is smaller or comparable to the gas cooling or
recombination times, this would be reflected in the radial profile of
the gas temperature and ionized fraction. Also, a strong jump in both
temperature and ionized fraction can be produced across a He II
ionization front. Even after the radiation source is dead, the
difference in the ionized fraction inside and outside a former He III
region would persist for quite a long time. 

Figure \ref{HII1} shows the evolution of the 21--cm
signal from the recombining H II region, produced at $z=20$ by a short-lived
source whose spectrum  is similar to a black body with a surface
temperature of $90,000$ K. Initially the gas brightness grows almost
linearly with the hydrogen neutral fraction. Subsequently, when the
ionized fraction falls below $\sim 0.2$, the intensity of Ly$\alpha$
photons (which  dominate over collisional coupling for $x$ above $\sim
0.05$) produced by recombinations is insufficient to fully decouple
$T_s$ from $T_{CMB}$, and, unless additional Ly$\alpha$
photons are provided by exterior radiation sources, the signal may drop.

\section{Galactic winds}
Even though large numbers of Ly$\alpha$ photons may be produced by the overdense gas in galaxies via recombination or excitation of hydrogen atoms, 
these photons do not typically affect the 21 cm emission in the neighbouring IGM, as they redshift out of resonance close to their 
origin. This limitation, however, may be bypassed if the peculiar velocity of the gas is very large. Ly$\alpha$ photons emitted by a 
medium moving with peculiar velocity $V$ may reenter the resonance within $R=V/H$ from the point of origin. 
Large peculiar velocities involving a large mass of gas may indeed be
generated by one or more supernovae explosions \cite{OC,KY,Wi}.

A wind expanding with velocity $V_s$ that exceeds 100 km/s shockheats
and compresses the surrounding gas into a thin shell. The  cooling in
the outer parts of the shell, where the temperature is above $\sim
10^6$ K, is  dominated by inverse Compton scatterings and
bremsstrahlung. In the inner parts, where the temperature drops below
$\sim 10^5$ K,  most of the energy is radiated away in Ly$\alpha$
photons.  By considering a thin shell expanding with velocity $V_s$ at
a radius $R_s$,
we find the distribution of the photons that are at resonance in the
local gas frame to be 
\begin{eqnarray}
\label{Lw2}
J_0=\left(\frac{L_\alpha H^{-1}c}{16\pi^3\nu_\alpha R^3}\right)
\left[\left(1-\frac{R_s^2}{R^2}\right)\left(\frac{V_s^2}{H^2R^2}-1\right)\right]^{-1/2} . 
\end{eqnarray}
Figure 1c shows the 21 cm signal produced by a galactic wind
for different radiation intensities. 

Apart from supernovae, thin shells expanding with large peculiar
velocities can be produced by large voids forming from steep initial
density perturbations. The distribution of Ly$\alpha$ photons produced
by such shells can still be described by equation \ref{Lw2}. However,
in this case the ratio $V_s/HR_s$  (which for shells driven by
supernovae can be very large) is typically between $1.2$ and $1.5$
\cite{CN}. 

\begin{figure}[t]
\resizebox{\columnwidth}{!}
{\includegraphics{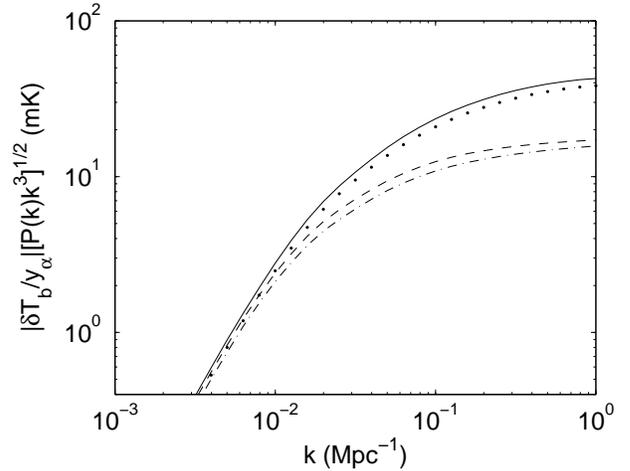}}
\caption{\label{spec}Power spectrum of the 21 cm signal at $z=20$. The pumping radiation is assumed to be produced by X-ray sources in halos with $T_{\rm vir}>10^4$ K or $T_{\rm vir}>5\cdot 10^3$ K (solid and dotted lines), or by Pop III stars in halos with $T_{\rm vir}>10^4$ K or $T_{\rm vir}>5\cdot 10^3$ K (dashed and dash-dotted lines). We assumed $y_\alpha\ll 5$.}
\end{figure}
\section{Short-lived radiation sources}
In the examples given in the previous sections, we have neglected the
possible time-dependence of the luminosity of radiation
sources. However, if the luminosity changes over a relatively short
time-scale, this will be reflected in the 21-cm signal. Since the
speed of light is finite, regions that are observed simultaneously
are pumped by photons emitted at different times. Therefore, by
measuring the 21--cm signal from different regions around the source,
it is possible to reconstruct the evolution of the source spectrum and
intensity. Figure \ref{Lymtd} illustrates the evolution of the 21-cm
profile after a starburst that lasts a few million years. 

\section{Summary}
We have shown that different radiation sources produce distinctly
different 21-cm profiles, which can be sensitive not only to the
source spectrum, but also to its history (\S 2 and 6). Furthermore,
differences can also be identified in the combined signal from
multiple radiation sources of different types. Figure \ref{spec} shows
the power spectrum of the 21-cm signal produced by either the X-ray or
the UV sources contributed by a biased distribution of halos (the
power spectrum calculation is based on the formalism developed by
Barkana \& Loeb (2005a)). Predictably, the X-ray sources, which
typically influence smaller regions, produce greater fluctuations on
small scales than do Pop III UV sources. This difference between X-ray and UV can be even larger 
if Pop III is replaced with Pop II stars, which produce lower proportion of higher n resonance photons (Barkana \& Loeb 2005b; Pritchard \& Furlanetto 2006).

Assuming that at $z\sim 20$ the typical distance between the first radiation 
sources is a few Mpc, their individual detection requires an angular resolution of 
about one arcminute (or baseline of $\sim$~15~km) and frequency resolution of $\sim 100$~kHz.
For a beamwidth and bandwidth just small enough to resolve such
objects, a detection of an X-ray source with $L\sim 10^{42}$ erg/s (which equals the 
luminosity of one or more black holes of total mass $\sim 10^4 M_\odot$ shining with Eddington luminosity) 
requires a sensitivity of $A/T\sim 10^3$~m$^2$K$^{-1}/\sqrt{(t_{\rm obs}/10^3 hours)}$, where
$t_{\rm obs}$ is the integration time. Thus it appears that SKA, whose planned sensitivity is 
 a few times $10^3$~m$^2$K$^{-1}$ at the relevant frequencies, would be able to detect such X-ray 
sources individually. 
The detection of the fluctuating 21-cm background produced by the superposition
of the first sources is even more promising. For example, the curves in Figure \ref{spec}
can be easily distinguished at scales $k\sim 10^{-2}$~Mpc$^{-1}$ by
LOFAR, SKA, or PAST for 4 weeks of integration time (see Zaldarriaga,
Furlanetto, \& Hernquist 2004 for details on power spectrum error
estimation).

\acknowledgments

This work was supported by a W.J. McDonald
Fellowship to L.~C., a Department of
Energy Computational Science Graduate Fellowship to  M.~A.~A., and NASA Astrophysical Theory Program grants
NAG5-10825 and NNG04G177G.

\end{document}